\documentclass[aps, prb, reprint, twocolumn, longbibliography, floatfix]{revtex4-2}
\usepackage[T1]{fontenc}           
\usepackage[british]{babel} 
\usepackage{mathtools}
\usepackage{graphicx} % Required for inserting images
\usepackage{hyperref}

\usepackage{color}

\renewcommand{\vec}[1]{\mathbf{#1}}

\newcommand{\grad}{\boldsymbol{\nabla}}
\newcommand{\ep}{\epsilon}
\newcommand{\vep}{\varepsilon}

\begin{document}

\title{Strong coupling polarons in cavity quantum materials:\\ limits of cavity-induced electron mass renormalization}

\author{D. M. Basko}
\affiliation{Univ. Grenoble Alpes, CNRS, LPMMC, 38000 Grenoble, France}
%\date{\today}

\begin{abstract}
Cavity-induced electron mass renormalization is an important ingredient of many proposals to control material properties through vacuum field modification. What is its ultimate limit? Here this question is answered by (i)~noting that this mass renormalization is essentially due to the field-mediated interaction between the electron and the polarizable degrees of freedom in the cavity mirrors, and (ii)~directly evaluating the corresponding Feynman path integral for a 2D electron by numerically exact quantum Monte Carlo. The sought upper limit is provided by the Landau-Pekar polaron in the planar geometry. 
The presented calculation bridges the gap between cavity control and dielectric engineering of material properties.
\end{abstract}

\maketitle

\section{Introduction}

%Possibility to control material properties by modifying vacuum fluctuations of quantized electromagnetic field via optical cavity engineering has generated significant interest in recent years. Renormalization of electron mass due to interaction with the vacuum field is a paradigmatic effect of quantum electrodynamics; its modification in the presence of a cavity was argued to be an important ingredient for the cavity engineering of material properties. It was predicted that cavity-induced changes in the mass could even reach values comparable to the mass itself. Then, a natural question arises: what are the fundamental limits for the cavity-induced mass renormalization? Its investigation is the subject of the present paper.

%The ability to control material properties through the engineering of optical cavities—by modifying the vacuum fluctuations of the quantized electromagnetic field—has generated significant interest in recent years. A paradigmatic example of quantum electrodynamics is the renormalization of electron mass due to its interaction with the vacuum field. It has been argued that the presence of a cavity can modify this effect, thereby playing a crucial role in the cavity-based engineering of material properties. Notably, cavity-induced changes in electron mass have been shown to reach values comparable to the mass itself. This raises a fundamental question: What are the ultimate limits of cavity-induced mass renormalization?

The possibility of controlling material properties by modifying vacuum fluctuations of quantized electromagnetic field via optical cavity engineering has generated significant interest (see Refs.~\cite{GarciaVidal2021, Schlawin2022, Mandal2023, Ruggenthaler2023, Lu2025, Baydin2025, Bretscher2026} for recent reviews).
Renormalization of electron mass due to interaction with the vacuum field is a paradigmatic effect in quantum electrodynamics (QED)~\cite{LandauLifshitzQED, PeskinSchroeder};
its modification in the presence of a cavity~\cite{Matloob2011} was argued to be an important ingredient for the cavity-based engineering of material properties~\cite{ Sentef2018, Rokaj2022, Mochida2024, Lu2024, Welakuh2025, Yang2026a, Fan2026, Yang2026b}.
Notably, cavity-induced changes in electron mass have been predicted to reach values comparable to the mass itself~\cite{Eckhardt2025}. This raises a fundamental question: what are the ultimate limits of cavity-induced mass renormalization?
This is the subject of the present paper.

%In this paper, this question is addressed based on the following key observation:
The key observation helping to address this question is the following.
The cavity-induced correction to the electron mass, when non-negligible, is mainly due to the Coulomb interaction between the electron and the polarizable degrees of freedom inside the cavity mirrors, as can be traced from the calculations of Ref.~\cite{Eckhardt2025}. Thus, the cavity-induced mass renormalization is just the conventional polaronic effect introduced a long time ago~\cite{Landau1933, Pekar1946, Landau1948, Lee1953, Gurari1953, Frohlich1954, Feynman1955, Devreese2003}: the field produced by the electron deforms the polarizable lattice, and the resulting polarization cloud follows the moving electron contributing to its inertia. The cavity QED setting thus corresponds to a geometry where the electron is spatially separated from the polarizable medium~\cite{Hipolito1979, deBodas1983, Sols1987, Mogulkoc2016}. 
The highest values of the polaron effective mass are obtained in the strong-coupling regime, as described by Landau and Pekar for a bulk crystal~\cite{Landau1948}; adapted to the present geometry, their picture provides the answer to the above question.
It should also be noted that sensitivity of material properties to dielectric environment, based on the original idea by Keldysh~\cite{Keldysh1979}, was observed in many experiments on 2D materials~\cite{Faugeras2015, Stier2016, Raja2017, Park2018, Waldecker2019, Kim2020, Peimyoo2020, Senger2021, Tebbe2023}. The present work aims at bridging the gap between cavity control and dielectric engineering of material properties.

\begin{figure}
\includegraphics[width=0.48\textwidth]{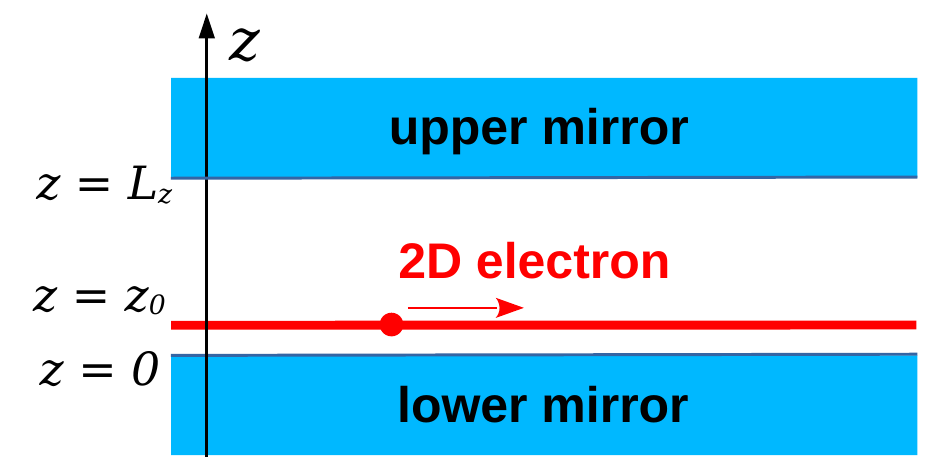}
\caption{\label{fig:cavity}
The studied system: a planar cavity with a 2D monolayer inside. The monolayer ($z=z_0$) is much closer to the lower mirror ($z=0$) than to the upper one ($z=L_z\gg{z}_0$), so electrostatic coupling to the upper mirror can be neglected.
}
\end{figure}

Here, I study the mass renormalization for a single electron in a 2D monolayer placed inside a planar cavity whose mirrors are made of a dielectric hosting polar optical phonons (Fig.~\ref{fig:cavity}). 
The interaction between the electron and the mirrors is taken in the electrostatic limit (known to dominate cavity QED at short distances~\cite{SaezBlazquez2023, Andolina2024, Pantazopoulos2024, SanchezMartinez2024, Riolo2025, Andolina2026, Andolina2025b}). 
This problem has been addressed in Ref.~\cite{deBodas1983} by a variational method with a surprising conclusion about a first-order transition, later disputed in Refs.~\cite{Xiaogang1985, Gerlach1987}.
Here I calculate the mass renormalization by a (numerically exact) quantum Monte-Carlo (QMC) evaluation of the Feynman path integral and compare it with the weak-coupling and strong-coupling calculations. While the former recovers the result of Ref.~\cite{Eckhardt2025}, the latter provides the sought upper limit for the cavity-induced mass renormalization in the considered system; it turns out to be of the same order as the bulk Landau-Pekar result~\cite{Landau1948}.
The found continuous dependence of the effective mass on the system's parameters excludes the possibility of a first-order transition, thus closing the debate~\cite{deBodas1983, Xiaogang1985, Gerlach1987}.

\section{The model}

The following model is chosen to explicitly include the phonon degrees of freedom. Then, it is checked to reproduce the standard description of dielectric mirrors forming the cavity.

Consider a planar cavity whose mirrors, occupying the half-spaces $z<0$ and $z>L_z$, are made of a material with the dielectric function
\begin{equation}\label{eq:vep=}
\tilde\vep(\omega) = \tilde\vep_\infty\,\frac{\omega_\text{L}^2 - \omega^2}{\omega_\text{T}^2-\omega^2},
\end{equation}
corresponding to the isotropic Lorentz oscillator model for polar optical phonons. Here $\omega_\text{T}$ is the mechanical frequency of the phonons (the bare vibration frequency determined by the ions' restoring forces alone), and $\omega_\text{L}$ is the Coulomb frequency including the extra contribution coming from the long-range electric field created by the polar displacement of longitudinal optical phonons. The high-frequency asymptotic value $\tilde\vep_\infty$ originates from polar excitations at higher energies. The cavity's interior is filled with a material whose dielectric constant~$\vep$ is taken to be frequency-independent. The plane $z=z_0$ contains a 2D monolayer hosting electrons with mass $m$ and parabolic dispersion $p^2/(2m)$, where $\vec{p}$ is the in-plane momentum. We assume the monolayer to be much closer to one of the mirrors ($z_0\ll{L}_z$), so in the following we just ignore the upper mirror (formally sending $L_z\to\infty$).

The electrons are described by a pair of Grassmann fields $\psi(\vec{r}_\|)$, $\psi^*(\vec{r}_\|)$ (in the following we will focus on a single electron, so the fermionic nature is not important, as well as the omitted spin index), and the phonons by the corresponding lattice displacement field $\vec{u}(\vec{r})$ [with the notations $\vec{r}\equiv(x,y,z)$, $\vec{r}_\|\equiv(x,y)$].
The lattice displacement is characterized by the reduced mass density~$\mu$; it is also associated with electric polarization in the material, $\vec{P}(\vec{r}) = \rho\vec{u}(\vec{r})$. The coefficient $\rho$ has the dimensionality of charge density and is determined by the microscopic charge distribution around the displaced ions; such local relation is valid under the same assumptions as neglecting the spatial dispersion in the dielectric function~(\ref{eq:vep=}), that is, at length scales exceeding the atomic scale.
The coupling to the electromagnetic field is taken in the electrostatic (Coulomb) limit, so the field is described by the electrostatic potential~$\varphi(\vec{r})$.

The zero-temperature action for this system, corresponding to the weight $e^{i\mathcal{S}}$ (in Gaussian units with $\hbar=1$), is given by:
\begin{subequations}\label{eqs:action}\begin{align}
&\mathcal{S}=\int{dt}\left[\int{d}^2\vec{r}_\|\,\mathcal{L}_\psi  + \int_{z<0}{d}^3\vec{r}\,\mathcal{L}_\vec{u} + \int{d}^3\vec{r}\,\mathcal{L}_\varphi\right],\\
&\mathcal{L}_\psi = \psi^*(\vec{r}_\|)\left[i\,\frac\partial{\partial{t}} + \frac{1}{2m}\nabla_\|^2- e\varphi(\vec{r}_\|,z_0)\right]\psi(\vec{r}_\|),\\
&\mathcal{L}_\vec{u} = \frac\mu2\left|\frac{\partial\vec{u}(\vec{r})}{\partial{t}}\right|^2
-\frac{\mu\omega_\text{T}^2}2|\vec{u}(\vec{r})|^2
-\grad\varphi(\vec{r})\cdot\rho\vec{u}(\vec{r}),\\
&\mathcal{L}_\varphi = \frac{\tilde{\vep}_\infty\,\theta(-z)+\vep\,\theta(z)}{8\pi}\,|\grad\varphi(\vec{r})|^2,
\end{align}\end{subequations}
where $\theta(z)$ is the Heaviside step function and we suppressed the time arguments of the fields for brevity. 
The Lagrangian $\mathcal{L}_\psi$ describes 2D electrons subject to the electrostatic potential at $z=z_0$. The three terms in $\mathcal{L}_\vec{u}$ represent, respectively, the kinetic energy of the lattice, the mechanical elastic energy, and the interaction of the lattice polarization $\rho\vec{u}(\vec{r})$ with the electric field $-\grad\varphi(\vec{r})$. The Lagrangian $-\mathcal{L}_\varphi$ is just the energy density of the electrostatic field.
The background dielectric constants $\vep$ and~$\tilde\vep_\infty$ can be viewed as the result of integrating out high-frequency degrees of freedom. Their effect is assumed to be included into the electron mass~$m$, as well as the effect of interaction with polarizable degrees of freedom of the monolayer itself~\cite{Sio2023, Shahnazaryan2025, Kudlis2026}.

To check that the action~(\ref{eqs:action}) indeed reproduces the dielectric function~(\ref{eq:vep=}), it is sufficient to consider the classical equations of motion for the lattice displacement and the potential in the frequency representation:
\begin{subequations}\label{eqs:uvarphirho}\begin{align}
&\mu(\omega_\text{T}^2-\omega^2)\vec{u}(\vec{r}) = -\rho\grad\varphi(\vec{r}),
\label{eq:u=}\\
&-\grad\cdot[\tilde{\vep}_\infty\,\theta(-z)+\vep\,\theta(z)]\grad\varphi(\vec{r}) = 4\pi\varrho(\vec{r}),
\label{eq:phi=}\\
&\varrho(\vec{r}) %= \varrho_\text{e}(\vec{r}) + \varrho_\text{ph}(\vec{r})
= e|\psi(\vec{r}_\|)|^2\delta(z-z_0)-\grad\cdot[\theta(-z)\rho\vec{u}(\vec{r})].
\label{eq:rho=}
\end{align}\end{subequations}
%Here the charge density $\varrho(\vec{r})$ has the electronic contribution $\varrho_\text{e}(\vec{r})=e|\psi(\vec{r}_\|)|^2\delta(z-z_0)$ and the phonon contribution $\varrho_\text{ph}(\vec{r})=-\grad\cdot[\theta(-z)\rho\vec{u}(\vec{r})]$. 
Expressing $\vec{u}(\vec{r})$ from Eq.~(\ref{eq:u=}) and plugging it into~(\ref{eq:rho=}) and~(\ref{eq:phi=}), one recovers the Poisson equation for $\varphi(\vec{r})$ with $e|\psi(\vec{r}_\|)|^2\delta(z-z_0)$ as the external source, and with the dielectric function $\tilde{\vep}(\omega)\,\theta(-z)+\vep\,\theta(z)$, thereby identifying
\begin{equation}
\omega_\text{L}^2-\omega_\text{T}^2 = \frac{4\pi\rho^2}{\tilde\vep_\infty\mu}.
\end{equation}

\section{Weak-coupling calculation}

It is instructive to start with the standard perturbative calculation. First, it elucidates the relation between the electron mass renormalization in cavity QED and electron-phonon polaron physics. Second, it actually covers many realistic situations, as will be seen later.
In fact, the same expressions for the binding energy and the effective mass are obtained from a more general variational approach where each phonon mode is assumed to be in some coherent state~\cite{Lee1953, Gurari1953, deBodas1983}, so they have a broader range of validity than just lowest-order expansion.

\begin{figure}
\includegraphics[width=0.48\textwidth]{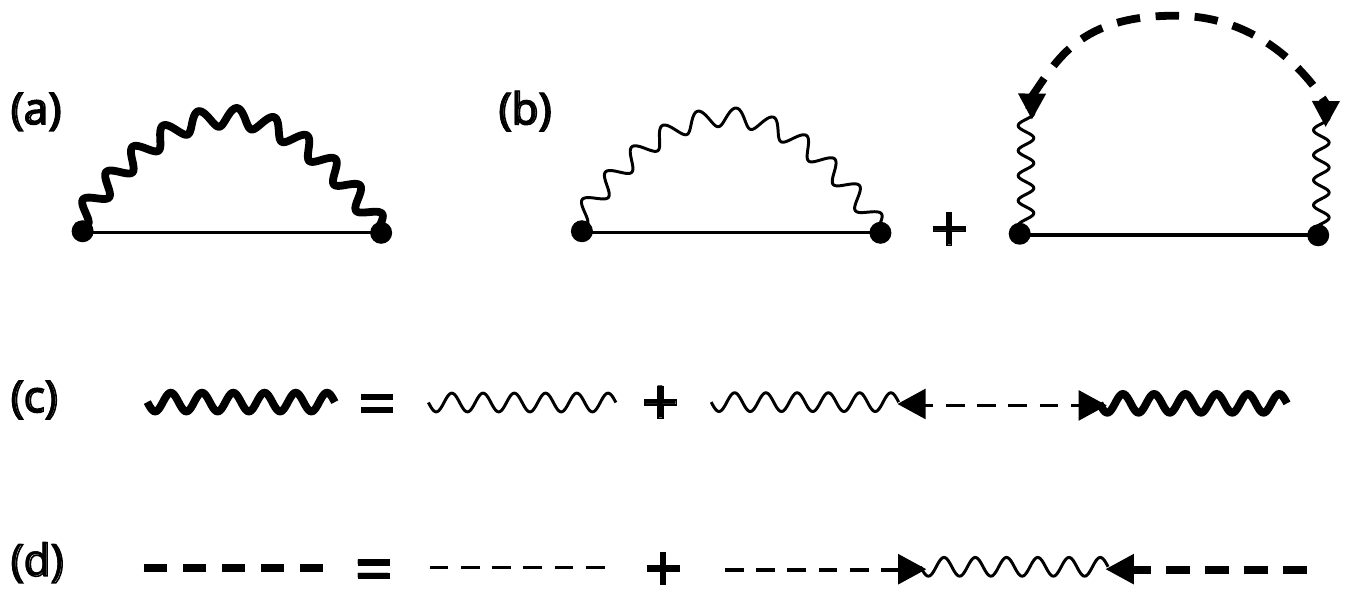}
\caption{\label{fig:diagrams}
(a), (b) Equivalent diagrams for the second-order electron self-energy. (c),~(d)~Dyson equations for the phonon-dressed  Coulomb field propagator $\tilde{V}(\vec{r},\vec{r}',\omega)$ (thick wavy line) and for the Coulomb-dressed phonon propagator $\tilde{D}_{ij}(\vec{r},\vec{r}',\omega)$ (thick dashed line), respectively. The thin lines represent the corresponding bare propagators, and the solid line represents the electron propagator. The circles and the triangles represent the electron charge vertex $e\delta(z-z_0)$, and the lattice charge vertex $\rho\grad$, respectively.
}
\end{figure}

The lowest-order (second order in the electron charge) electronic self-energy $\Sigma(\vec{p},\ep)$ can be calculated using the standard diagrammatic approach~\cite{NegeleOrland}. It can be represented diagrammatically in two equivalent ways (shown in Fig.~\ref{fig:diagrams}(a) and (b), respectively): either via the propagator $\tilde{V}(\vec{r},\vec{r}',\omega)$ of the Coulomb field~$\varphi$, dressed by the phonons, or via the propagator $\tilde{D}_{ij}(\vec{r},\vec{r}',\omega)$ of the phonon field~$\vec{u}$, dressed by the Coulomb interaction ($i,j=x,y,z$ label the Cartesian components); in the latter case, the bare Coulomb propagator $V(\vec{r},\vec{r}')$ determines the electron-phonon coupling. The resulting expansion of $\Sigma(\vec{p},\ep)$ on the mass shell $\ep=p^2/(2m)$ at small momenta,
\begin{equation}
\Sigma(\vec{p},p^2/(2m)) = \Sigma_0 + \Sigma_2\,\frac{p^2}{2m} + O(p^4),
\end{equation}
gives the polaron binding energy $-\Sigma_0$ and the relative mass increase $\Delta{m}/m=-\Sigma_2$.

The phonon-dressed Coulomb field propagator can be found from the Poisson equation
\begin{equation}\label{eq:Poisson}
-\grad\cdot[\tilde\vep(\omega)\,\theta(-z)+\vep\,\theta(z)]\grad\tilde{V}(\vec{r},\vec{r}',\omega) = 4\pi\delta(\vec{r}-\vec{r}'),
\end{equation}
obtained by integrating out the phonon field from action~(\ref{eqs:action}). Due to the in-plane translation invariance, it is natural to make the Fourier transform with respect to $\vec{r}_\|-\vec{r}_\|'$ introducing the in-plane wave vector~$\vec{k}$. Then the solution of Eq.~(\ref{eq:Poisson}) for $z,z'>0$ reads
\begin{align}\label{eq:tildeV=}
&\tilde{V}(\vec{k},z,z',\omega) = \frac{2\pi}{\vep{k}}\left[e^{-k|z-z'|} + \frac{\vep-\tilde\vep(\omega)}{\vep+\tilde\vep(\omega)}\,e^{-k(z+z')}\right]
\end{align}
and has poles $\omega=\pm(\omega_\text{s}-{i}0^+)$ at the surface phonon frequency, $\omega_\text{s}^2 = {(\vep\omega_\text{T}^2+\tilde{\vep}_\infty\omega_\text{L}^2)/(\vep+\tilde{\vep}_\infty)}$. The second-order electron self-energy,
\begin{align}\label{eq:Sigma=V}
\Sigma(\vec{p},\ep) = {}&{} i\int\frac{d\omega}{2\pi}\,\frac{d^2\vec{k}}{(2\pi)^2}\,\frac{\tilde{V}(\vec{k},z_0,z_0,\omega)}{\ep-\omega-|\vec{p}-\vec{k}|^2/(2m)+i0^+},
\end{align}
is determined by the pole at $\omega_\text{s}-{i}0^+$. Its small-momentum expansion at $\ep=\ep_\vec{p}\equiv p^2/(2m)$ gives
%\begin{align}
%-\Sigma(\vec{p},\ep_\vec{p}) = {}&{} \alpha_\text{s}
%\int\limits_0^\infty{dx}\,e^{-2\zeta{x}}\left[\frac{\omega_\text{s}}{x^2+1} + \frac{2\ep_\vec{p}x^2}{(x^2+1)^3}+\ldots\right]
%% {}\nonumber\\ = {}&{} 
%%\alpha_\text{s}\begin{cases}
%%\omega_\text{s} + \ep_\vec{p}/4 + O(p^4), & \zeta\ll1,\\
%%\omega_\text{s}/(\pi\zeta) + \ep_\vec{p}/(\pi\zeta^3) + O(p^4), & \zeta\gg1,
%%\end{cases}
% {}\nonumber\\ = {}&{} 
%\alpha_\text{s}\begin{cases}
%\pi\omega_\text{s}/2 + \pi\ep_\vec{p}/8 + O(p^4), & \zeta\ll1,\\
%\omega_\text{s}/(2\zeta) + \ep_\vec{p}/(2\zeta^3) + O(p^4), & \zeta\gg1,
%\end{cases}
%\label{eq:Sigma2=}
%\end{align}
\begin{equation}
\frac{\Delta{m}}m = \alpha_\text{s}
\int\limits_0^\infty{d\xi}\, \frac{2\xi^2e^{-2\zeta\xi}}{(\xi^2+1)^3}
= \begin{cases}
\alpha_\text{s}\pi/8, & \zeta\ll1,\\
\alpha_\text{s}/(2\zeta^3), & \zeta\gg1,
\end{cases}
\label{eq:Sigma2=}
\end{equation}
where we introduced the dimensionless electron-phonon coupling strength~$\alpha_\text{s}$ and distance to the mirror~$\zeta$:
\begin{equation}
\alpha_\text{s} = \frac{2\tilde\vep_\infty(\omega_\text{L}^2-\omega_\text{T}^2)}{(\vep+\tilde\vep_\infty)^2\omega_\text{s}^2}\,
\sqrt{\frac{me^4}{2\omega_\text{s}}},\quad
\zeta = z_0\sqrt{2m\omega_\text{s}}.
\end{equation}
The $\zeta\gg1$ asymptotics~(\ref{eq:Sigma2=}) reproduces Eq.~(12) of Ref.~\cite{Eckhardt2025}, thereby confirming the equivalence of the two calculations in this regime. If one wants to relax the Coulomb approximation adopted here, the structure of the perturbation theory and the diagrams shown in Fig.~\ref{fig:diagrams} remain the same; one just should use the full field propagators found from the Maxwell's equations, and include vector coupling vertices.

\section{Monte-Carlo calculation}

The polaron problem can be handled in a numerically exact and unbiased way by QMC~\cite{Prokofiev1998, Mishchenko2000, Titantah2001, Luo2025}. To apply Markov-chain QMC, action~(\ref{eqs:action}), suitable for perturbative calculation, can be equivalently transformed by (i)~passing to the first quantization for a single electron, (ii)~integrating out exactly the harmonic degrees of freedom~$\varphi$ and~$\vec{u}$, and (iii)~passing to imaginary time~$\tau$ on a circle of circumference~$\beta$, the inverse temperature~\cite{Feynman1955}. The resulting Euclidean action reads
\begin{align}
S_\text{E} = {}&{}
\int_0^\beta{d}\tau\,\frac{m}2\left|\frac{d\vec{r}_\|(\tau)}{d\tau}\right|^2
\nonumber \\ & {}
+ \frac{e^2}2\int_0^\beta{d}\tau\,d\tau'\,\tilde{V}_\text{l}(\vec{r}_\|(\tau)-\vec{r}_\|(\tau'),\tau-\tau'),
\label{eq:SE=}
\end{align}
where the lattice-mediated self-attraction of the electron is given by the Matsubara-frequency ($\omega_n\equiv{2}\pi{n}/\beta$) Fourier transform of the lattice-induced part of Eq.~(\ref{eq:tildeV=}):
\begin{align}
\tilde{V}_\text{l}(\vec{r}_\|,\tau) = {}&{} 
%\int\frac{d^2\vec{k}}{(2\pi)^2}\sum_{\omega_n}
%\frac{e^{i\vec{k}\vec{r}_\|-i\omega_n\tau}}\beta
%\frac{2\pi}{\vep{k}}\,\frac{\vep-\tilde\vep(i\omega_n)}{\vep+\tilde\vep(i\omega_n)}\,e^{-2kz_0}
\int\frac{d^2\vec{k}}{(2\pi)^2}\sum_{n=-\infty}^\infty
\frac{e^{i\vec{k}\vec{r}_\|-i\omega_n\tau}}\beta\nonumber\\
{}&{}\times\left[\tilde{V}(\vec{k},z_0,z_0,i\omega_n)-\tilde{V}(\vec{k},z_0,z_0,\infty)\right]
 {}\nonumber\\
= {}&{}
-\frac{\alpha_\text{s}\sqrt{2\omega_\text{s}/(me^4)}}{\sqrt{|\vec{r}_\||^2+4z_0^2}}\,
\frac{\omega_\text{s}\cosh(\beta\omega_\text{s}/2-\omega_\text{s}|\tau|)}{2\sinh(\beta\omega_\text{s}/2)}
\nonumber \\ \equiv {}&{} V_\text{eff}(\vec{r}_\|)\,\Upsilon(\tau).
\label{eq:tildeVtau=}
\end{align}
Rescaling lengths and energies by $z_0$ and $\omega_\text{s}$ leaves only two dimensionless parameters in the problem: $\alpha_\text{s}$ and~$\zeta$.

The polaron mass $m+\Delta{m}$ can be extracted by noting that at $|\tau-\tau'|\to\infty$, the probability distribution of $|\vec{r}_\|(\tau)-\vec{r}_\|(\tau')|$ is dominated by the softest degrees of freedom corresponding to the free translational motion of a composite particle, and thus should tend to a Gaussian corresponding to a free particle with a modified mass~\cite{Feynman1955}.
Treating $e^{-S_\text{E}}$ as a probability distribution for periodic trajectories $\vec{r}_\|(\tau)=\vec{r}_\|(\tau+\beta)$, we calculate the average
\begin{equation}\label{eq:Wbeta=}
W(\beta) = \left\langle\int_0^\beta\frac{d\tau\,d\tau'}{\beta^3}|\vec{r}_\|(\tau)-\vec{r}_\|(\tau')|^2\right\rangle
\mathop{\to}\limits_{\beta\to\infty}\frac{1/3}{m+\Delta{m}}.
\end{equation}
The $\tau$~integration is introduced to reduce statistical error, and the factor 1/3 is found by  a free-particle calculation. 

$W(\beta)$ is calculated by discretizing the time $\tau_j=j\delta$, $j=1,\ldots,N$, $\delta=\beta/N$, in Eqs.~(\ref{eq:SE=})--(\ref{eq:Wbeta=}) and evaluating the resulting $2N$-dimensional integral by Markov-chain Monte-Carlo~\cite{Ceperley1995}. A simple direct scheme is adopted here: exact sampling of the kinetic term in $e^{-S_\text{E}}$ by a Brownian bridge is followed by Metropolis acceptance or rejection according to the interaction term. For each set of parameters, a fixed bridge length was chosen to keep the acceptance rate $\sim0.2-0.4$. For a fixed~$\delta$, $W(\beta)$ was calculated for several values of~$\beta$ and extrapolated to $\beta\to\infty$ by a linear fit in $1/\beta$,
%\footnote{The polynomial can be taken of first or second order.Corrections beyond $1/\beta^2$ are probably exponential in~$\beta$; this can be shown explicitly for the Feynman's model action of two masses coupled by an elastic spring, and is also expected to hold for the polaron whose internal degrees of freedom are gapped.} 
checking convergence at small~$\delta$ (see Appendix~\ref{app:QMC} for details).

\section{Landau-Pekar approach}

The central object of the Landau-Pekar polaron theory is the single-electron wave function $\Psi(\vec{r}_\|)$ of the bound state in the static potential well created by the lattice displacements. It represents a static solution of the classical equations of motion obtained by replacing the Grassmann variables $\psi(\vec{r}_\|)\to\Psi(\vec{r}_\|)$ in the action~(\ref{eqs:action}), corresponding to the minimum of its potential energy part. Upon elimination of the variables $\varphi(\vec{r})$ and $\vec{u}(\vec{r})$ using Eqs.~(\ref{eqs:uvarphirho}) at $\omega=0$, the wave function $\Psi(\vec{r}_\|)$ is found as the minimizer of the static Pekar's functional~\cite{Pekar1946},
\begin{align}
F[\Psi] = {}&{}\frac1{2m}\int{d^2}\vec{r}_\|\,|\grad_\|\Psi(\vec{r}_\|)|^2 \nonumber\\
{}&{} + \frac{e^2}2\int{d^2}\vec{r}_\|\,{d^2}\vec{r}_\|^\prime\,
V_\text{eff}(\vec{r}_\|-\vec{r}_\|')
|\Psi(\vec{r}_\|)|^2|\Psi(\vec{r}_\|')|^2,
\label{eq:Pekar=}
\end{align}
under the constraint $\int|\Psi(\vec{r}_\|)|^2\,{d}^2\vec{r}_\|=1$.
Here $V_\text{eff}(\vec{r}_\|)$ is the same as in Eq.~(\ref{eq:tildeVtau=}). More details are given in Appendix~\ref{app:Landau-Pekar}.

This approach, based on a classical treatment of the lattice, is justified when the polarization cloud contains many phonons, that is, the polaron binding energy $|\min{F}[\Psi]|\gg\omega_\text{s}$. The same condition, read as the typical electronic time scale being much shorter than the lattice scale, ensures that the lattice effectively sees the average electron charge density $e|\Psi(\vec{r}_\|)|^2\delta(z-z_0)$. Functional~(\ref{eq:Pekar=}) can also be obtained from the quantum Hamiltonian using a many-body variational wave function in the form of a direct product $|\Psi\rangle\otimes|\Phi_\text{ph}\rangle$ with the electron state $|\Psi\rangle$ determined by the wave function $\Psi(\vec{r}_\|)$ and $|\Phi_\text{ph}\rangle$ being a bosonic coherent state whose parameters are to be optimized (see Appendix~\ref{app:Pekar}).

While the static functional determines the polaron binding energy, its effective mass is found by constructing a solution of the classical equations of motion for action~(\ref{eqs:action}), which moves with a constant velocity~$\vec{v}$, that is, $\vec{u}(\vec{r},t) = \vec{u}_\vec{v}(\vec{r}_\|-\vec{v}t,z)$,  $\varphi(\vec{r},t) = \varphi_\vec{v}(\vec{r}_\|-\vec{v}t,z)$, and $\Psi(\vec{r}_\|,t) = e^{im\vec{v}\vec{r}_\|-i(mv^2/2)t-i\ep{t}}\Psi_\vec{v}(\vec{r}_\|-\vec{v}t)$. Then, the energy should be evaluated on this solution and expanded to the order $O(v^2)$.
As argued in Ref.~\cite{Landau1948} in the bulk case, one can then neglect the deformation of the static solution by a finite velocity [also $O(v^2)$], and since the static solution minimizes the potential energy, the only correction to the energy comes from the lattice kinetic energy; this argument remains valid here as well (see  Appendix~\ref{app:effective_mass}). Thus, we have
\begin{equation}
\frac{\Delta{m}\,v^2}2 = \int_{z<0}{d}^3\vec{r}\,\frac\mu2\left|\frac{\partial\vec{u}(\vec{r}_\|-\vec{v}t,z)}{\partial{t}}\right|^2,
\end{equation}
where $\vec{u}(\vec{r})$ is found from Eqs.~(\ref{eqs:uvarphirho}) at $\omega=0$ in terms of the static solution $\Psi(\vec{r}_\|)$.

\section{Discussion of results}

\begin{figure}
\includegraphics[width=0.48\textwidth]{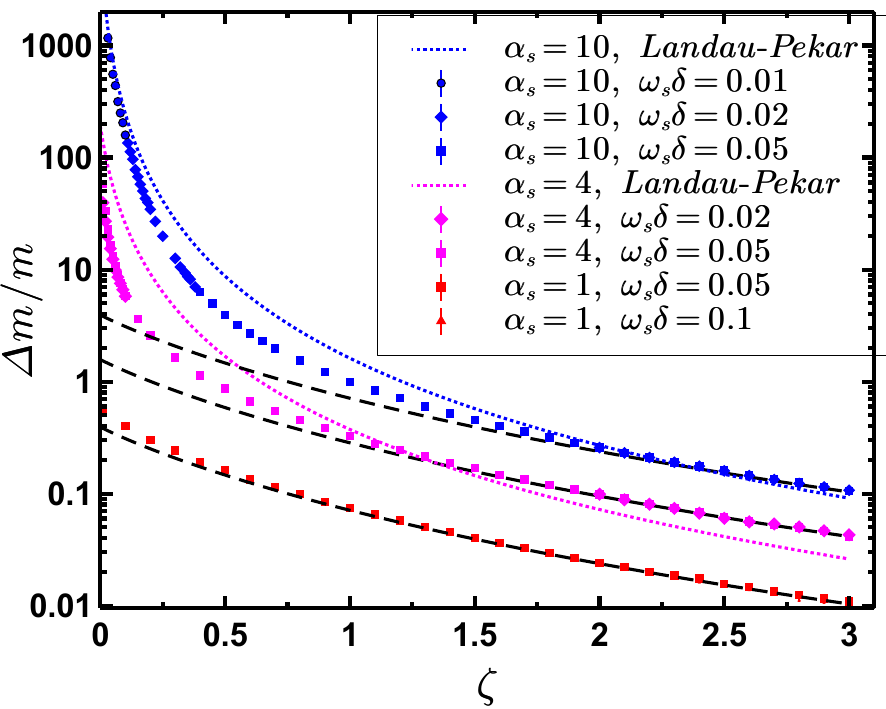}
\caption{\label{fig:mass_vs_zeta}
Relative correction to the electron mass as a function of the dimensionless distance $\zeta\equiv\sqrt{2m\omega_\text{s}}z_0$ for three values of the coupling constant $\alpha_\text{s} = 1,\,4,\,10$. Dashed lines, symbols and dotted lines show the perturbative result~(\ref{eq:Sigma2=}), the QMC result, and the result of the Landau-Pekar approach, respectively. The QMC error bars are within the symbol size.
}
\end{figure}

The results of all calculations are presented in Fig.~\ref{fig:mass_vs_zeta}. As expected, the weak coupling expression~(\ref{eq:Sigma2=}) works at small~$\alpha_\text{s}$ or large~$\zeta$. At large~$\alpha_\text{s}$ and small~$\zeta$, the QMC results approach those of the Landau-Pekar calculation. The $\zeta\to0$ limit of the latter, $\Delta{m}/m=C\alpha_\text{s}^4$ with $C=0.732\ldots$ (see Appendix~\ref{app:LPnumerics}), at $\alpha_\text{s}\gtrsim1$ provides the sought upper limit for the cavity-induced electron mass renormalization, the main result of the present work.

The putative first-order transition, found in Ref.~\cite{deBodas1983} and supposed to  be accompanied by a large jump in the polaron mass, would be expected at $\zeta\approx0.05$ for $\alpha_\text{s}=4$ and at $\zeta\approx0.56$ for $\alpha_\text{s}=10$. The QMC results smoothly connect the strong-coupling and the weak coupling limits without any sign of a discontinuity, which was probably an artefact of the variational ansatz used in Ref.~\cite{deBodas1983}.

Strong-coupling limit is not easy to reach in practice.
One example is SrTiO${}_3$ with $\tilde{\vep}_\infty=5$, $\omega_\text{L}=170\:\mbox{cm}^{-1}$~\cite{Servoin1980, Kamaras1995} and $\omega_\text{T}$~strongly temperature-dependent, softening down to $10\:\mbox{cm}^{-1}$ at low temperatures~\cite{Inoue1981}. Taking $\vep=4$ (SiO$_2$) and $m=m_0$, the free electron mass, yields $\alpha_\text{s}=6.5$; however, this material is quite exceptional precisely because of the TO phonon softening, while more typical values are lower. $\alpha_\text{s}\gg1$ might be achievable in Moiré structures with narrow electronic bands ($m\gg{m}_0$).

The results of this paper can be applied to metallic mirrors: Drude metal corresponds to setting $\tilde{\vep}_\infty=1$, $\omega_\text{T}\to0$ and $\omega_\text{L}=\omega_\text{p}$ (the plasma frequency) in Eq.~(\ref{eq:vep=}). Then the coupling constant is given by $\alpha_\text{s}=(\vep+1)^{-3/4}\sqrt{2me^4/\omega_\text{p}}$. For $\omega_\text{p} = 8.5\:\text{eV}$ (gold~\cite{Olmon2012}), $\vep=2$ (SiO${}_2$ at frequency $\omega_\text{s}\approx5\:\mbox{eV}$) and $m=m_0$ this gives $\alpha_\text{s}=1.1$.

The characteristic length scale $(2m\omega_\text{s})^{-1/2}$ adimensionalizing the separation~$z_0$ turns out to be $1.5\:\mbox{nm}$ for SrTiO${}_3$ and $1\:\mbox{\AA}$ for gold. At such short distances the mirror material cannot be described by a macroscopic dielectric function without spatial dispersion; the sharp rise of $\Delta{m}/m$ at $\zeta\to0$ and $\alpha_\text{s}>1$ will be effectively cut off at the atomic scale, where the problem should be handled by a microscopic calculation~\cite{Dai2026}. The macroscopic approach adopted here is expected to work at distances exceeding a few nanometers; then, from Fig.~\ref{fig:mass_vs_zeta} we see that at $\zeta>2$ the weak-coupling expression~(\ref{eq:Sigma2=}) works reasonably even for $\alpha_\text{s}=10$.

The electrostatic approximation, adopted here, breaks down at large distances $z_0\sim{c}/(\sqrt{\vep}\omega_\text{s})$ ($c$~being the speed of light), when retardation effects become important. This corresponds to $\sim6\:\mu\mbox{m}$ for SrTiO${}_3$ and $\sim30\:\mbox{nm}$ for gold. While in the former case the second cavity mirror can be closer, the latter scale is quite relevant. Still, note that at this crossover scale the mass renormalization itself is already negligible: $\Delta{m}/m\sim\alpha_\text{s}[\omega_\text{s}/(mc^2)]^{3/2}$. Thus, for all practical purposes, Coulomb coupling between the electron and polar excitations in cavity mirrors can be considered to be the only source of cavity-induced electron mass renormalization.

\section{Conclusions}

I presented a non-perturbative calculation of the mass renormalization for a single electron in a 2D monolayer placed inside a planar cavity with Drude-Lorentz dielectric mirrors in the electrostatic limit. Depending on the distance $z_0$ to the nearest mirror, the results smoothly interpolate between the perturbative calculation including retardation~\cite{Eckhardt2025} and the strong-coupling polaron limit~\cite{Landau1948} in the 2D geometry, thus bridging the gap between cavity control and dielectric engineering of material properties.

The strong-coupling $z_0\to0$ limit, $\Delta{m}/m\approx0.732\,\alpha_\text{s}^4$, establishes an upper bound for $\Delta{m}$ at $\alpha_\text{s}\gtrsim1$. However, this limit is hard to reach in realistic structures. In many cases, the weak-coupling Coulomb self-energy expression~(\ref{eq:Sigma2=}), easily generalizable to arbitrary dielectric environments, provides an adequate description of the effect.

\begin{acknowledgments}
I am grateful to L.~Herviou and M.~Holzmann for advice on numerics. The codes and the numerical data are publicly available~\cite{Zenodo}.
\end{acknowledgments}


\appendix

\section{Details on Monte-Carlo calculation}
\label{app:QMC}

For a given set of parameters (namely, the dimensionless coupling constant~$\alpha_\text{s}$, the dimensionless distance~$\zeta$, the dimensionless inverse temperature $\omega_\text{s}\beta$ and the dimensionless discretization step $\omega_\text{s}\delta\equiv \omega_\text{s}\beta/N$), a Markov chain in the $2N$-dimensional trajectory space $\underline{x}\equiv(x_1,y_1,\ldots,x_N,y_N)$ is constructed by a sequence of successive updates. Each update consists of constructing two independent Brownian bridges of length $\ell$ (this length is fixed for the whole QMC run), one for~$x$ and one for~$y$, on a segment $j=j_0+1,\ldots,j_0+\ell-1$ of the trajectory ($j_0$~is chosen randomly, and the trajectory is cyclic, $x_{j+N}\equiv{x}_j$, $y_{j+N}\equiv{y}_j$). That is, the values $x_{j_0}$~and~$x_{j_0+\ell}$ remain unchanged, while new values for $x_{j_0+1},\ldots,x_{j_0+\ell-1}$ are proposed, sampled exactly from the Gaussian probability distribution
\begin{align}
&P(x_{j_0+1},\ldots,x_{j_0+\ell-1})\nonumber\\
&{}\propto\exp\left[-\frac{(x_{j_0}-x_{j_0+1})^2}{2\delta/m}-\ldots-\frac{(x_{j_0+\ell-1}-x_{j_0+\ell})^2}{2\delta/m}\right].
\end{align} 
The same procedure is followed for $y_{j_0+1},\ldots,y_{j_0+\ell-1}$, thus yielding a new trajectory~$\underline{x}'$. This update is then accepted with probability $\min\{1, e^{S_\text{int}(\underline{x})-S_\text{int}(\underline{x}')}\}$, where the discretized version of the interaction action in Eq.~(\ref{eq:SE=}) is given by
\begin{subequations}\begin{align}
&S_\text{int}(\underline{x}) = -\sum_{i,j=1}^N\frac{\alpha_\text{s}(\omega_\text{s}\delta)^2\Upsilon_{ij}}{\sqrt{2m\omega_\text{s}[(x_i-x_j)^2 + (y_i-y_j)^2 + 4z_0^2]}},\\
&\Upsilon_{ij} \equiv \frac{\cosh(\beta\omega_\text{s}/2-\omega_\text{s}\delta|i-j|)}{2\sinh(\beta\omega_\text{s}/2)}.
\end{align}\end{subequations}
The Brownian bridge length~$\ell$ is chosen in order to keep the average acceptance rate between 0.2 and 0.4; for weak coupling it can be the whole trajectory, while for strong coupling it can be just a few sites thus requiring a long simulation to sample the configuration space. The Markov chain correlation length is estimated using a binning analysis for the observable $W(\beta)$ [Eq.~(\ref{eq:Wbeta=})]; this correlation length is then used to estimate the error bars on the observable. For each QMC run, a random initial condition is chosen by sampling the kinetic energy only, and then the Markov chain is allowed to relax before measuring the observable.

\begin{figure}
\includegraphics[width=0.48\textwidth]{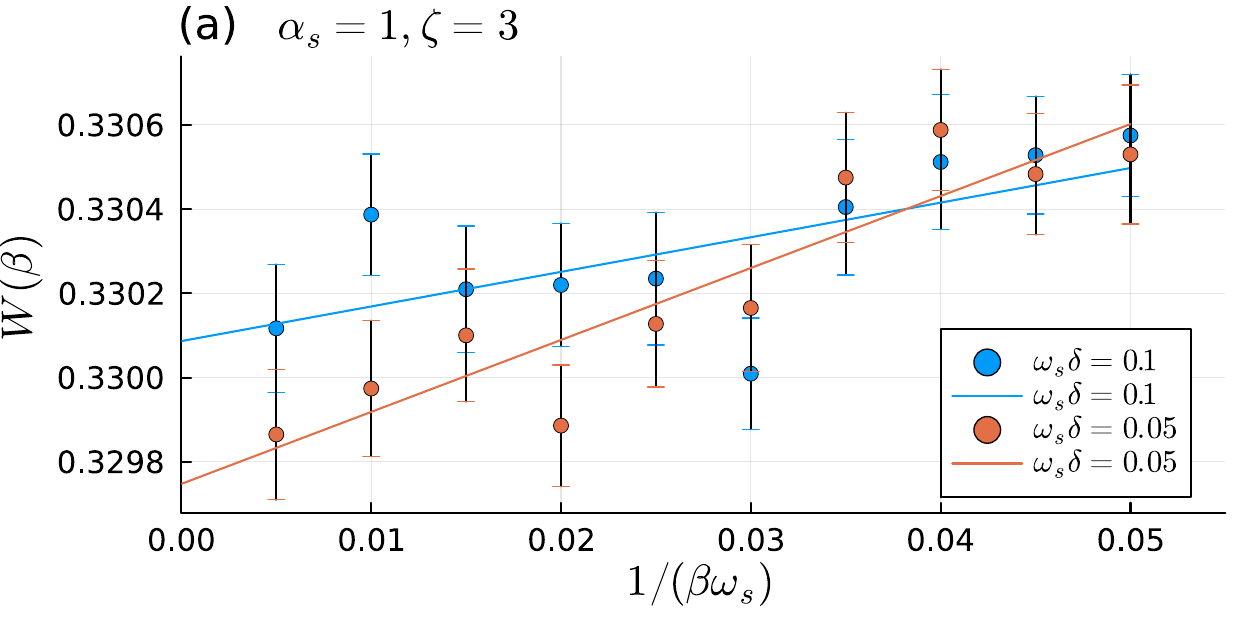}\vspace{1mm}\\
\includegraphics[width=0.48\textwidth]{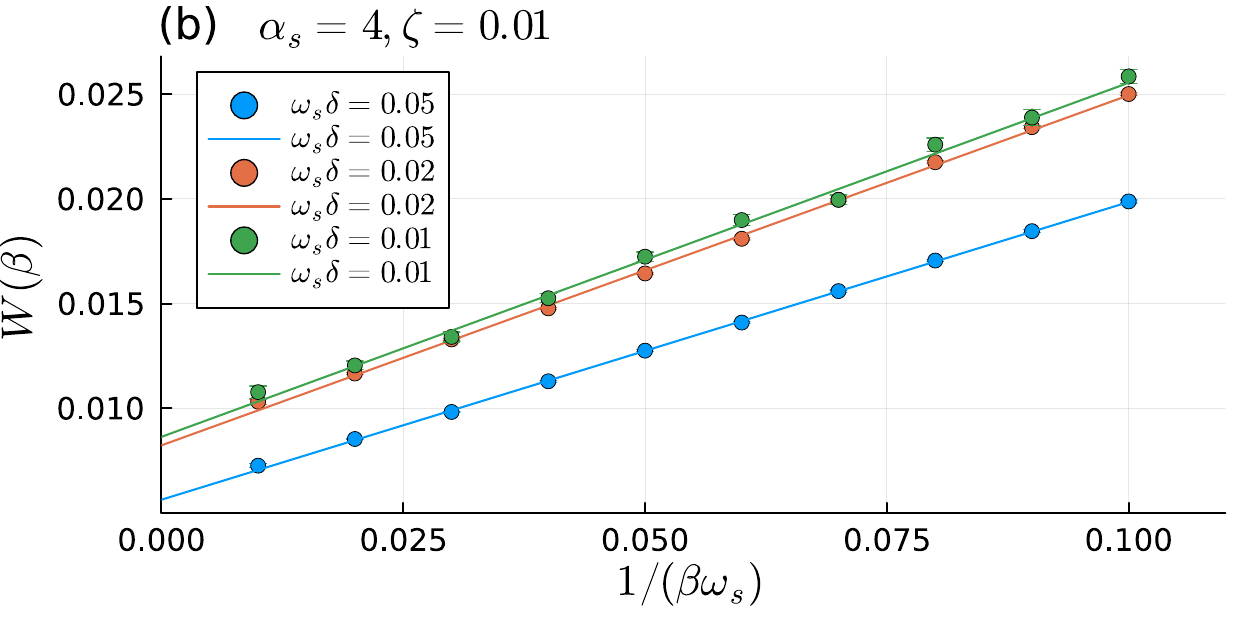}\vspace{1mm}\\
\includegraphics[width=0.48\textwidth]{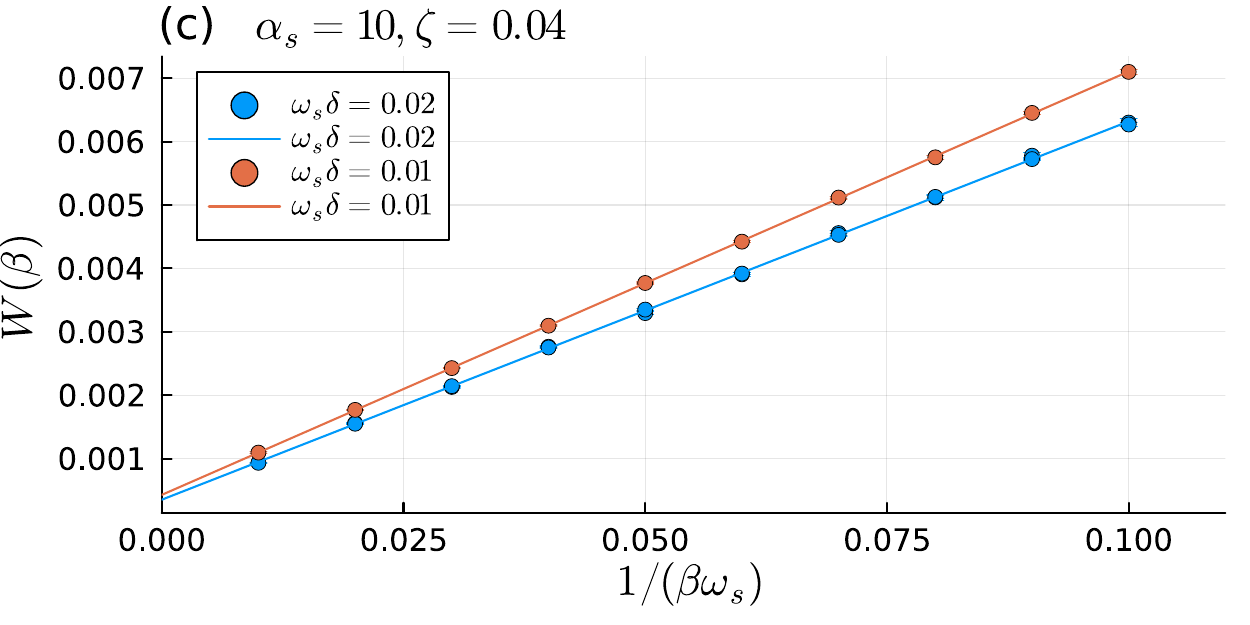}
\caption{\label{fig:QMCextrapolation}
$\beta\to\infty$ extrapolation for three points from Fig.~\ref{fig:mass_vs_zeta}.
}
\end{figure}
%\begin{figure*}
%\includegraphics[width=0.32\textwidth]{fig_QMCextrapolation_weak.pdf}
%\includegraphics[width=0.33\textwidth]{fig_QMCextrapolation_typical.pdf}
%\includegraphics[width=0.32\textwidth]{fig_QMCextrapolation_strong.pdf}
%\caption{\label{fig:QMCextrapolation}
%$\beta\to\infty$ extrapolation for three points from Fig.~\ref{fig:mass_vs_zeta}.
%}
%\end{figure*}

The numerical coefficient in Eq.~(\ref{eq:Wbeta=}) can be determined by calculating the average $\langle|x(\tau)-x(\tau')|^2\rangle$ for a free one-dimensional particle of mass $m_*$ with the action $\int_0^\beta(m_*/2)(dx/d\tau)^2\,d\tau$:
\begin{align}
\langle|x(\tau)-x(\tau')|^2\rangle = {}&{} \frac4\beta\sum_{n=1}^\infty\frac{1-\cos\omega_n(\tau-\tau')}{m_*\omega_n^2}\nonumber\\
= {}&{} \frac{|\tau-\tau'|(\beta-|\tau-\tau'|)}{\beta{m}_*}.
\label{eq:xtauxtaup2}
\end{align}
In $d$~dimensions, the result for $\langle|\vec{r}(\tau)-\vec{r}(\tau')|^2\rangle$ is just $d$~times larger since each Cartesian component enters independently in the observable and in the free action. Integration over $\tau,\tau'$ yields Eq.~(\ref{eq:Wbeta=}) for $d=2$.
For a polaron, Eq.~(\ref{eq:xtauxtaup2}) with $m_*=m+\Delta{m}$ is valid at large $|\tau-\tau'|$, well exceeding (i)~the inverse energy gap of the polaron's internal degrees of freedom, as well as (ii)~the inverse energy scale associated with non-parabolicity of the lowest energy band. Thus, the integral relation~(\ref{eq:Wbeta=}) is valid up to corrections $O(1/\beta)$.
%The same quantity can be easily evaluated for the Feynman's trial action describing a particle of mass~$m$ connected to another particle of mass~$M$ by an elastic spring with the spring constant $\Xi$, for which the denominator $m\omega_n^2$ in Eq.~(\ref{eq:xtauxtaup2}) should be replaced by $m\omega_n^2 + M\omega_\text{r}^2\omega_n^2/(\omega_\text{r}^2+\omega_n^2)$, where $\omega_\text{r}^2=\Xi(m+M)/(mM)$ corresponds to the oscillatory relative motion of the two masses. This gives
%\begin{equation}
%\frac{W(\beta)}{W(\infty)} = 1 + \frac{6M/m}{\beta\omega_\text{r}}
%\left(\coth\frac{\beta\omega_\text{r}}2 - \frac{2}{\beta\omega_\text{r}}\right).
%\end{equation}
%Note that the power series in $1/\beta$ here truncates at the second order, the remainder being exponentially small. It is plausible to conjecture this property to hold for a generic polaron whose internal degrees of freedom have an energy gap, without making assumptions on the coefficients. This inspires the $\beta\to\infty$ extrapolation of $W(\beta)$ by fitting it to a low-degree polynomial in $1/(\beta\omega_\text{s})$.

For fixed $\alpha_\text{s}$,~$\zeta$ and~$\omega_\text{s}\delta$, the observable $W(\beta)$ is calculated for a set of $\beta$~values (typically, for $1/(\beta\omega_\text{s})=0.01,\,0.02,\ldots,\,0.1$) and linearly extrapolated to $1/\beta\to0$.
This procedure is illustrated in Fig.~\ref{fig:QMCextrapolation} for two extreme points from Fig.~\ref{fig:mass_vs_zeta} with the weakest and the strongest coupling (panels (a) and (c), respectively) and a typical point [panel~(b)]. The main difficulty in the weak coupling case is that a relatively small correction~$\Delta{m}$ on top of the unperturbed value~$m$ has to be detected, so finite error bars lead to a significant loss of relative precision. 
At strong coupling, one needs larger~$\beta$ and smaller~$\delta$ (and thus, large~$N$), while the allowed Brownian bridge length~$\ell$ shrinks; this results in a long correlation length of the Markov chain ($\sim2\times10^6$ updates and $\ell=20$ for the leftmost point in Fig.~\ref{fig:mass_vs_zeta}) leading to long computational times. For typical points in Fig.~\ref{fig:mass_vs_zeta}, the extrapolation is quite reliable, as illustrated in Fig.~\ref{fig:QMCextrapolation}(b).

\section{Details on the Landau-Pekar approach}
\label{app:Landau-Pekar}

\subsection{Hamiltonian}
\label{app:Hamiltonian}

To pass to the Hamiltonian formulation of the theory, one can integrate out the scalar potential $\varphi$~from the action~(\ref{eqs:action}) and perform the canonical quantization. This results in the following Hamiltonian:
\begin{widetext}\begin{align}\label{eq:Hpsiu}
\hat{H} = {}&{} \int{d}^2\vec{r}_\|\,\hat\psi^\dagger(\vec{r}_\|)\left(-\frac{1}{2m}\nabla_\|^2
\right)\hat\psi(\vec{r}_\|) + \int{d^3}\vec{r}\,d^3\vec{r}'\,e\hat\psi^\dagger(\vec{r}_\|)\,\hat\psi(\vec{r}_\|)\,V(\vec{r}_\|,z_0,\vec{r}')\,\rho[-\grad'\cdot\theta(-z')\hat{\vec{u}}(\vec{r}')]\nonumber\\
{}&{}+\int_{z<0}d^3\vec{r}\left[\frac{1}{2\mu}|\hat{\boldsymbol{\pi}}(\vec{r})|^2 + \frac{\mu\omega_\text{T}^2}2|\hat{\vec{u}}(\vec{r})|^2\right] + \frac{\rho^2}2\int{d^3}\vec{r}\,d^3\vec{r}'\,V(\vec{r},\vec{r}')\,
[\grad\cdot\theta(-z)\hat{\vec{u}}(\vec{r})][\grad'\cdot\theta(-z')\hat{\vec{u}}(\vec{r}')].
\end{align}\end{widetext}
%\begin{align}
%\hat{H} = {}&{} \int{d}^2\vec{r}_\|\,\hat\psi^\dagger(\vec{r}_\|)\left(-\frac{1}{2m}\nabla_\|^2\right)\hat\psi(\vec{r}_\|) \nonumber\\ {}&{} + \int{d^3}\vec{r}\,d^3\vec{r}'\,e\hat\psi^\dagger(\vec{r}_\|)\,\hat\psi(\vec{r}_\|)\,V(\vec{r}_\|,z_0,\vec{r}')\,\rho[-\grad'\cdot\theta(-z')\hat{\vec{u}}(\vec{r}')]\nonumber\\
%{}&{}+\int_{z<0}d^3\vec{r}\left[\frac{1}{2\mu}|\hat{\boldsymbol{\pi}}(\vec{r})|^2 + \frac{\mu\omega_\text{T}^2}2|\hat{\vec{u}}(\vec{r})|^2\right]\nonumber\\ {}&{} + \frac{\rho^2}2\int{d^3}\vec{r}\,d^3\vec{r}'\,V(\vec{r},\vec{r}')\,
%[\grad\cdot\theta(-z)\hat{\vec{u}}(\vec{r})][\grad'\cdot\theta(-z')\hat{\vec{u}}(\vec{r}')].\label{eq:Hpsiu}
%\end{align}
Here $\hat\psi^\dagger(\vec{r}_\|)$ and $\hat\psi(\vec{r}_\|)$ are the fermionic creation and annihilation operators for the electrons, while $\hat{\vec{u}}(\vec{r})$ and $\hat{\boldsymbol{\pi}}(\vec{r})$ are the operators of the phonon displacement and its conjugate momentum density, respectively. The bare Coulomb propagator $V(\vec{r},\vec{r}')$ is the Green's function of the Poisson equation~(\ref{eq:Poisson}) with $\tilde\vep(\omega)$ replaced by the background dielectric constant~$\tilde{\vep}_\infty$.

Next, one should find the phonon normal modes which diagonalize the second line in Eq.~(\ref{eq:Hpsiu}) including the long-range Coulomb interaction (the Coulomb phonons). We do not need their explicit form; we just use the in-plane translational invariance of the system to conclude that these modes can be labeled by the in-plane wave vector~$\vec{k}$, and denote all other quantum numbers by~$\lambda$ which encodes the polarization, the $z$~dependence, and the surface or bulk nature of these modes. These modes have some frequencies $\omega_{\vec{k},\lambda}$ and are described by the phonon creation and annihilation operators~$\hat{b}_{\vec{k},\lambda}^\dagger$ and~$\hat{b}_{\vec{k},\lambda}$.

For a single electron, it is convenient to pass to the first quantization. Then the Hamiltonian becomes
\begin{align}
\hat{H} = {}&{} -\frac{\nabla_\|^2}{2m} + \sum_{\vec{k},\lambda} \left(g_{\vec{k},\lambda} \hat{b}_{\vec{k},\lambda}e^{i\vec{k}\vec{r}_\|} + g_{\vec{k},\lambda}^* \hat{b}_{\vec{k},\lambda}^\dagger e^{-i\vec{k}\vec{r}_\|}\right)
{}\nonumber\\  {}&{}+ \sum_{\vec{k},\lambda}\omega_{\vec{k},\lambda}\left(\hat{b}_{\vec{k},\lambda}^\dagger\hat{b}_{\vec{k},\lambda}+\frac12\right).
\label{eq:Hrb}
\end{align}
Here the coupling constants $g_{\vec{k},\lambda}$ are obtained by writing the electron-phonon coupling term [the second term in Eq.~(\ref{eq:Hpsiu})] in the normal-mode basis; their explicit form will never be needed. The total momentum of the system, conserved by Hamiltonian~(\ref{eq:Hrb}), has the form
\begin{equation}\label{eq:ptot=}
\hat{\vec{p}}_\text{tot} = -i\grad_\| + \sum_{\vec{k},\lambda}\vec{k}\hat{b}_{\vec{k},\lambda}^\dagger\hat{b}_{\vec{k},\lambda}.
\end{equation}

\subsection{Pekar's functional}
\label{app:Pekar}

Pekar's functional can be obtained from the separable variational state $|\mathcal{X}\rangle$ with an electronic wave function $\Psi(\vec{r}_\|)$ and the phonons in a coherent state
\begin{equation}\label{eq:Psir=}
\langle\vec{r}_\||\mathcal{X}\rangle = \Psi(\vec{r}_\|)\prod_{\vec{k},\lambda}\exp \left[\chi_{\vec{k},\lambda} \hat{b}_{\vec{k},\lambda}^\dagger - \chi^*_{\vec{k},\lambda} \hat{b}_{\vec{k},\lambda}\right]|0\rangle.
\end{equation}
The expectation value of Hamiltonian~(\ref{eq:Hpsiu}) in this state can be straightforwardly related to the functional~(\ref{eq:Pekar=}). Indeed, keeping in mind the normal mode expansion of the lattice displacement and the momentum density,
\begin{subequations}\label{eqs:upi=b}\begin{align}
& \hat{\vec{u}}(\vec{r}) = \sum_{\vec{k},\lambda} \sqrt{\frac1{2\mu\omega_{\vec{k},\lambda}}}\,\frac{e^{i\vec{k}\vec{r}_\|}}{\sqrt{S}}\, \vec{w}_{\vec{k},\lambda}(z)\,\hat{b}_{\vec{k},\lambda} + \mbox{h.c.}, \\
& \hat{\boldsymbol{\pi}}(\vec{r}) = -i\sum_{\vec{k},\lambda}\sqrt{\frac{\mu\omega_{\vec{k},\lambda}}2} \frac{e^{i\vec{k}\vec{r}_\|}}{\sqrt{S}}\, \vec{w}_{\vec{k},\lambda}(z)\,\hat{b}_{\vec{k},\lambda} + \mbox{h.c.},
\end{align}\end{subequations}
with some normalized mode functions $\vec{w}_{\vec{k},\lambda}(z)$,
\begin{equation}
\int_{-\infty}^0\vec{w}_{\vec{k},\lambda}^*(z)\cdot\vec{w}_{\vec{k},\lambda'}(z)\,dz = \delta_{\lambda\lambda'},
\end{equation}
one can note the following.
\begin{itemize}
\item[(i)]
Dropping the constant zero-point energy term in the phonon Hamiltonian is equivalent to imposing normal ordering of the phonon creation and annihilation operators in the Hamiltonian~(\ref{eq:Hpsiu}) with the substitution~(\ref{eqs:upi=b}).
\item[(ii)]
Since bosonic coherent states are eigenstates of the annihilation operator, averaging the normal-ordered Hamiltonian~(\ref{eq:Hpsiu}) over the coherent state~(\ref{eq:Psir=}) amounts to replacing $\hat{b}_{\vec{k},\lambda}\to\chi_{\vec{k},\lambda}$,  $\hat{b}_{\vec{k},\lambda}^\dagger \to \chi^*_{\vec{k},\lambda}$ in Eqs.~(\ref{eqs:upi=b}). These, in turn, substitute the operator fields $\hat{\boldsymbol{\pi}}(\vec{r})$, $\hat{\vec{u}}(\vec{r})$ in Eq.~(\ref{eq:Hpsiu}) by the classical ones.
\item[(iii)]
Since $e^{i\vec{k}\vec{r}_\|}\vec{w}_{\vec{k},\lambda}(z)$ form a complete functional basis in the half-space $z<0$, minimization with respect to the complex variational parameters $\chi_{\vec{k},\lambda},\chi_{\vec{k},\lambda}^*$ is equivalent to minimization with respect to the pair of classical real fields $\boldsymbol{\pi}(\vec{r})$ and $\vec{u}(\vec{r})$. While the former leads to $\boldsymbol{\pi}(\vec{r})=0$, the latter yields Eqs.~(\ref{eqs:uvarphirho}) with $\omega=0$ and the electronic charge density $e|\Psi(\vec{r}_\|)|^2\delta(z-z_0)$; substitution of the minimizing $\vec{u}(\vec{r})$ into the Hamiltonian yields the static Pekar's functional~(\ref{eq:Pekar=}).
\end{itemize}
Note that the state~(\ref{eq:Psir=}) is not an eigenstate of the total momentum~(\ref{eq:ptot=}).

For a real cylindrically symmetric function $\Psi({r}_\|)$, we perform the angular integration analytically,
\begin{align}
&\int_0^{2\pi}\frac{d\phi}{\sqrt{r_1^2+r_2^2+2r_1r_2\cos\phi+4z_0^2}}
\nonumber\\
{}&{}=\frac{4}{\sqrt{(r_1+r_2)^2+4z_0^2}}\,K\!\left(\frac{4r_1r_2}{(r_1+r_2)^2+4z_0^2}\right),
\end{align}
where $K(m)$ is the complete elliptic integral. 
We also change the variables $s=r_\|^2/z_0^2$, $\Psi(r_\|) = f(s)/z_0$, which gives the following functional to minimize under the constraint $\pi\int_0^\infty{f}^2(s)\,ds=1$:
\begin{align}
\frac{\zeta^2\,F[f]}{4\pi\omega_\text{s}} = {}&{} \int_0^\infty\left[\frac{df(s)}{ds}\right]^2s\,ds \nonumber\\ {}&{} - \frac{\alpha_\text{s}\zeta}2\int_0^\infty\frac{f^2(s)\,f^2(s')\,ds\,ds'}{\sqrt{(\sqrt{s}+\sqrt{s'})^2+4}}\,
\nonumber\\ {}&{}\qquad\times K\!\left(\frac{4\sqrt{ss'}}{(\sqrt{s}+\sqrt{s'})^2+4}\right).
\end{align}
The corresponding nonlinear Schr\"odinger equation reads
\begin{align}
\frac{\zeta^2\ep}{4\omega_\text{s}}\,{f}(s)
= {}&{} -\frac{d}{ds}\left[s\,\frac{d{f}(s)}{ds}\right] \nonumber \\
{}&{} - \alpha_\text{s}\zeta\int_0^\infty\frac{f^2(s')\,ds'}{\sqrt{(\sqrt{s}+\sqrt{s'})^2+4}}\,\nonumber\\ {}&{} \qquad\times K\!\left(\frac{4\sqrt{ss'}}{(\sqrt{s}+\sqrt{s'})^2+4}\right)f(s).
\label{eq:nonlinearSchroedinger}
\end{align}
The minimization is performed numerically by putting~$s$ on a grid and iterating the discrete eigenvalue problem.
Note that the rescaled functional depends on a single combination $\alpha_\text{s}\zeta$ of the two dimensionless parameters.

\subsection{Effective mass correction}
\label{app:effective_mass}

Semiclassically, the electronic wave function, the lattice displacement field, and the electrostatic potentials produced by the electron and the lattice satisfy the following system of equations:
\begin{subequations}\label{eqs:LandauPekar}\begin{align}
& i\,\frac{\partial\Psi(\vec{r}_\|,t)}{\partial{t}} = \left[-\frac{1}{2m}\,\nabla_\|^2 + e\varphi^\text{(l)}(\vec{r}_\|,z_0,t)\right]\Psi(\vec{r}_\|,t),\\
& \mu\,\frac{\partial^2\vec{u}(\vec{r},t)}{\partial{t}^2} = -\mu\omega_\text{T}^2\vec{u}(\vec{r},t) - \rho\grad[\varphi^\text{(e)}(\vec{r},t)+\varphi^\text{(l)}(\vec{r},t)],\\
&-\grad\cdot[\vep\,\theta(z)+\tilde{\vep}_\infty\theta(-z)]\grad\varphi^\text{(e,l)}(\vec{r},t) = 4\pi\varrho^\text{(e,l)}(\vec{r},t)).
\end{align}\end{subequations}
where the electron and lattice charge densities are given by 
\begin{subequations}\begin{align}
&\varrho^\text{(e)}(\vec{r},t)\equiv e|\Psi(\vec{r}_\|,t)|^2\delta(z-z_0),\\
&\varrho^\text{(l)}(\vec{r},t)\equiv -\rho\grad\cdot\theta(-z)\,\vec{u}(\vec{r},t).
\end{align}\end{subequations}
The solutions can be sought in the form
\begin{align*}
&\vec{u}(\vec{r},t) = \vec{u}_{\vec{v}}(\vec{r}_\|-\vec{v}t,z),\\
&\varphi^\text{(e,l)}(\vec{r}_\|,z,t) = \varphi_{\vec{v}}^\text{(e,l)}(\vec{r}_\|-\vec{v}t,z),\\
&\Psi(\vec{r}_\|,t) = e^{im\vec{v}\vec{r}_\|-i(mv^2/2)t-i\ep{t}}\Psi_{\vec{v}}(\vec{r}_\|-\vec{v}t).
\end{align*}
The corresponding energy is given by
\begin{align}
E = {}&{} \frac{mv^2}{2}+\int{d}^2\vec{r}_\|\,\frac{\left|\grad_\|\Psi(\vec{r}_\|)\right|^2}{2m} + {}\nonumber\\ {}&{} + 
\int{d}^3\vec{r}\,\theta(-z)\left[\frac\mu2\left|\frac{\partial\vec{u}(\vec{r})}{\partial{t}}\right|^2 + \frac{\mu\omega_\text{T}^2}2|\vec{u}(\vec{r})|^2\right]
+ {}\nonumber\\ {}&{} + 
\frac12\int{d}^2\vec{r}\,d^2\vec{r}'\, \varrho^\text{(l)}(\vec{r})\, V(\vec{r},\vec{r}')\, \varrho^\text{(l)}(\vec{r}')
+ {}\nonumber\\ {}&{} + 
\int{d}^2\vec{r}_\|\,d^3\vec{r}'\, e|\Psi(\vec{r}_\|)|^2 V(\vec{r}_\|,z_0,\vec{r}')\,\varrho^\text{(l)}(\vec{r}').
\label{eq:LandauPekarEnergy}
\end{align}
At $\vec{v}=0$, the static solution of Eqs.~(\ref{eqs:LandauPekar}) reproduces the minimizer of Pekar's functional~(\ref{eq:Pekar=}). Indeed, substituting $\vec{u}(\vec{r})$ from the second equation into the third one, one finds
\begin{subequations}\begin{align}
&\varphi^\text{(e)}(\vec{r}) + \varphi^\text{(l)}(\vec{r}) = \int{d}^2\vec{r}_\|'\,\tilde{V}(\vec{r},\vec{r}_\|',z_0,\omega\to0)\,e|\Psi(\vec{r}_\|')|^2,\\
&\varphi^\text{(e)}(\vec{r}) = \int{d}^2\vec{r}_\|'\,V(\vec{r},\vec{r}_\|',z_0)\,e|\Psi(\vec{r}_\|')|^2.
\end{align}\end{subequations}
The static solution of Eqs.~(\ref{eqs:LandauPekar}) also reproduces the minimizer of the static part of the functional~(\ref{eq:LandauPekarEnergy}). Finally, minimization of the latter with respect to~$\vec{u}$ yields Eq.~(\ref{eq:Pekar=}). 

At finite $\vec{v}$, the static solutions are deformed, the correction being $O(v^2)$. Since the static solution is the minimizer of the static part of the functional~(\ref{eq:LandauPekarEnergy}), the first variation vanishes. Thus, the $O(v^2)$ correction to the energy comes only from the lattice kinetic energy~\cite{Landau1948}:
\begin{align}
\Delta{E}={}&{}\int\limits_{z<0}{d}^3\vec{r}\,\frac\mu2\left|\frac{\partial\vec{u}(\vec{r})}{\partial{t}}\right|^2 {}\nonumber\\
={}&{} \frac{\rho^2}{2\mu\omega_\text{T}^4}\int_{z<0^-}{d}^3\vec{r}\,\left|(\vec{v}\cdot\grad)\grad\varphi^\text{tot}(\vec{r})\right|^2 {}\nonumber\\
={}&{} \frac{\rho^2v_iv_j}{2\mu\omega_\text{T}^4}\left[\int{d}^2\vec{r}_\|\,\frac{\partial\varphi^\text{tot}(\vec{r}_\|,0^-)}{\partial{x}_i}\,\frac{\partial^2\varphi^\text{tot}(\vec{r}_\|,0^-)}{\partial{x}_j\partial{z}}\right.{}\nonumber\\
{}&{} \qquad - \left. \int_{z<0^-}{d}^3\vec{r}\,\frac{\partial\varphi^\text{tot}(\vec{r})}{\partial{x}_i}\,\frac{\partial^3\varphi^\text{tot}(\vec{r})}{\partial{x}_j\partial{x}_k\partial{x}_k}\right],
\label{eq:DeltaE_byparts}
\end{align}
with the notation $\varphi^\text{tot}=\varphi^\text{(e)} + \varphi^\text{(l)}$.
Since
\begin{equation}
\varphi^\text{tot}(\vec{r}_\|,z<0) = \frac{2}{\vep+\tilde{\vep}(0)}\int\frac{d^2\vec{r}'\,e|\Psi(\vec{r}_\|')|^2}{\sqrt{|\vec{r}_\|-\vec{r}_\|'|^2 + (z-z_0)^2}},
\end{equation}
so $\nabla^2\varphi^\text{tot}(\vec{r}_\|,z<0)\propto\delta(z-z_0)$, and the second term in Eq.~(\ref{eq:DeltaE_byparts}) vanishes. The first term becomes
\begin{align}
\Delta{E}= {}&
\frac{\rho^2v^2}{2[\vep+\tilde{\vep}(0)]^2\mu\omega_\text{T}^4}\left(-\frac\partial{\partial{z}_0}\right)
\int{d}^2\vec{r}_\|\,{d}^2\vec{r}_\|'\,{d}^2\vec{r}_\|''\nonumber\\
{}&{}\times\frac{(\vec{r}_\|-\vec{r}_\|')\cdot(\vec{r}_\|-\vec{r}_\|'')\,e|\Psi(\vec{r}_\|')|^2\,e|\Psi(\vec{r}_\|'')|^2}{(|\vec{r}_\|-\vec{r}_\|'|^2+z_0^2)^{3/2}(|\vec{r}_\|-\vec{r}_\|''|^2+z_0^2)^{3/2}} = {}\nonumber\\
= {}& \frac{mv^2}{2}\,\frac{3\alpha_\text{s}z_0^4}{\pi\zeta^3}\int{d}^2\vec{r}_\|
\left[\int\frac{(\vec{r}_\|-\vec{r}_\|')\,|\Psi(\vec{r}_\|')|^2\,d^2\vec{r}_\|'}{(|\vec{r}_\|-\vec{r}_\|'|^2+z_0^2)^{3/2}}\right]
\nonumber\\
{}&{}\qquad \cdot\left[\int\frac{(\vec{r}_\|-\vec{r}_\|'')\,|\Psi(\vec{r}_\|'')|^2\,d^2\vec{r}_\|''}{(|\vec{r}_\|-\vec{r}_\|''|^2+z_0^2)^{5/2}}\right],
\end{align}
where the derivative $\partial/\partial{z}_0$ acts only on the explicit $z_0$ in the denominators; strictly speaking, the minimizer wave function $\Psi(\vec{r}_\|)$ also depends on~$z_0$, and $\partial/\partial{z}_0$ does not touch this dependence.
The angular integrals are expressed in terms of the complete elliptic integrals $K(m)$ and $E(m)$:
\begin{subequations}
\begin{align}
&\int\frac{(\vec{r}_\|-\vec{r}_\|')\,|\Psi({r}_\|')|^2\,d^2\vec{r}_\|'}{(|\vec{r}_\|-\vec{r}_\|'|^2+z_0^2)^{n/2}} \nonumber \\
&{} = \frac{\vec{r}_\|}{{r}_\|^2}\int_0^\infty|\Psi(r')|^2\,r'\,dr'
\nonumber\\ {}&{}\qquad\times\int_0^{2\pi}\frac{({r}_\|^2-{r}_\|r'\cos\phi)\,d\phi}{(z_0^2+{r}_\|^2+r'{}^2-2{r}_\|r'\cos\phi)^{n/2}} \equiv{}\nonumber\\
&{}\equiv \frac{\vec{r}_\|}{z_0^{n-2}r_\|^2}\int_0^\infty\frac{\mathcal{E}_{n/2}(r_\|,r')}{n/2-1}\,|\Psi(r')|^2\,r'\,dr',
\end{align}
\begin{align}
\mathcal{E}_{3/2}(r,r') ={}&{}
\frac{r^2-r'{}^2-z_0^2}{(r-r')^2+z_0^2}\,\frac{z_0}{\sqrt{(r+r')^2+z_0^2}}\,\nonumber\\
{}&{}\quad\times E\!\left(\frac{4rr'}{(r+r')^2+z_0^2}\right) 
{}\nonumber\\ {}&{} 
+ \frac{z_0}{\sqrt{(r+r')^2+z_0^2}}\,K\!\left(\frac{4rr'}{(r+r')^2+z_0^2}\right),
\end{align}
\begin{align}
\mathcal{E}_{5/2}(r,r') = {}&{} z_0^3\,\frac{7r^4-6r^2(r'{}^2-z_0^2)-(r'{}^2+z_0^2)^2}{[(r-r')^2+z_0^2]^2[(r+r')^2+z_0^2]^{3/2}}\,\nonumber\\ {}&{}\quad \times E\!\left(\frac{4rr'}{(r+r')^2+z_0^2}\right)  {}\nonumber\\ {}&{} 
+ z_0^3\,\frac{z_0^4-2z_0^2(r-r')r'-(r-r')^3(r+r')}{[(r-r')^2+z_0^2]^2[(r+r')^2+z_0^2]^{3/2}}\,\nonumber\\ {}&{}\quad \times K\!\left(\frac{4rr'}{(r+r')^2+z_0^2}\right).
\end{align}
\end{subequations}
In terms of the solution of Eq.~(\ref{eq:nonlinearSchroedinger}), this gives
\begin{align}
\frac{\Delta{m}}m = \frac{\alpha_\text{s}}{\zeta^3}\int_0^\infty\frac{ds}s
\int_0^\infty{d}s'\,\mathcal{E}_{3/2}(\sqrt{s}z_0,\sqrt{s'}z_0)\,f^2(s')\nonumber \\
{}\times\int_0^\infty{d}s''\,\mathcal{E}_{5/2}(\sqrt{s}z_0,\sqrt{s''}z_0)\,f^2(s'').
\label{eq:dmmE32E52}
\end{align}

\subsection{Simple variational functions}
\label{app:Gaussian}

A useful benchmark is provided by the Gaussian variational function, parametrized by its inverse spatial extent~$\kappa$:
\begin{equation}\label{eq:PsiGauss=}
\Psi(\vec{r}_\|) = \frac{\kappa}{\sqrt\pi}\,e^{-\kappa^2{r}_\|^2/2}.
\end{equation}
Using the Fourier components of $\Psi$ and $\Psi^2$,  
\begin{equation}
\Psi_\vec{k} = \frac{2\sqrt\pi}\kappa\,e^{-k^2/(2\kappa^2)},\quad
(\Psi^2)_\vec{k} = e^{-k^2/(4\kappa^2)},
\end{equation}
and the identities
\begin{align*}
&\int\frac{d^2\vec{r}_\|\,e^{-i\vec{k}\vec{r}_\|}}{\sqrt{r_\|^2+z^2}} = \frac{2\pi}k\,e^{-k|z|},\\
&\int\frac{d^2\vec{r}_\|\,\vec{r}_\|e^{-i\vec{k}\vec{r}_\|}}{(r_\|^2+z^2)^{3/2}} = -2\pi{i}\,\frac{\vec{k}}{k}\,e^{-k|z|},
\end{align*}
the energy functional can be evaluated analytically:
\begin{align}
\frac{\zeta^2F[\Psi]}{\omega_\text{s}} {}={} &  \int{d}^2\vec{r}_\|\,\left|\nabla_\|\Psi(\vec{r}_\|)\right|^2 \nonumber \\ {}&{} - \alpha_\text{s}\zeta\int{d}^2\vec{r}_\|\,d^2\vec{r}_\|'\,\frac{|\Psi(\vec{r}_\|)|^2|\Psi(\vec{r}_\|')|^2}{\sqrt{|\vec{r}_\|-\vec{r}_\|'|^2+4z_0^2}} = \nonumber\\
 = {}&{} \int\frac{d^2\mathbf{k}}{(2\pi)^2}\left[k^2|\Psi_{\mathbf{k}}|^2 - \alpha_\text{s}\zeta\,\frac{2\pi}k\,e^{-2kz_0}|(\Psi^2)_{\mathbf{k}}|^2\right]
 \nonumber\\ &  
 = \kappa^2 -\frac{\alpha_\text{s}\zeta}{2}\,\sqrt{2\pi}\kappa{e}^{2\kappa^2}\mathrm{erfc}(\sqrt{2}\kappa),
 \label{eq:FGauss=}
\end{align}
as well as the correction to the effective mass:
\begin{align}
\frac{\Delta{m}}m = {}&{} \frac{\alpha_\text{s}}{\pi\zeta^3}\int\frac{d^2\vec{k}}{(2\pi)^2} (2\pi)^2ke^{-2kz_0}|(\Psi^2)_{\mathbf{k}}|^2 \nonumber\\
= {}&{} \frac{\alpha_\text{s}}{\zeta^3}\left[\kappa^2(1+4\kappa^2)\sqrt{2\pi}\,\kappa{e}^{2\kappa^2}\mathrm{erfc}(\sqrt{2}\kappa) - 4\kappa^4\right].
\end{align}
Using the asymptotics
\begin{align}
&\sqrt{2\pi}\,\kappa{e}^{2\kappa^2}\mathrm{erfc}(\sqrt{2}\kappa) \nonumber\\
{}&{} = \begin{cases} \sqrt{2\pi}\kappa-4\kappa^2+O(\kappa^3),&\kappa\ll1,\\
1 - 1/(4\kappa^2) + 3/(16\kappa^4) + O(1/\kappa^6),&\kappa\gg1,\end{cases}
\end{align}
one can extract the asymptotic behaviour of various quantities:
\begin{align*}
\alpha_\text{s}\zeta\ll1: {}&{} \; \kappa_\text{min}= \sqrt{\frac\pi8}\, \alpha_\text{s}\zeta+ O((\alpha_\text{s}\zeta)^2),\\ {}&{} \; \frac{\zeta^2F_\text{min}}{\omega_\text{s}}=-\frac{\pi}{8}\,(\alpha_\text{s}\zeta)^2,\quad \frac{\Delta{m}}m=\frac{\pi^2}{16}\,\alpha_\text{s}^4,\\
\alpha_\text{s}\zeta\gg1: {}&{} \; \kappa_\text{min} = \left(\frac{\alpha_\text{s}\zeta}{8}\right)^{1/4} + O((\alpha_\text{s}\zeta)^{-1/4}),\\ {}&{} \; \frac{\zeta^2F_\text{min}}{\omega_\text{s}} = -\frac{\alpha_\text{s}\zeta}{2} + \sqrt{\frac{\alpha_\text{s}\zeta}{2}},\quad \frac{\Delta{m}}m=\frac{\alpha_\text{s}}{2\zeta^3}.
\end{align*}
In the latter limit $\alpha_\text{s}\zeta\gg1$, the Gaussian $\Psi(\vec{r}_\|)$ is expected to become the exact minimizer of the Pekar functional. Indeed, in this limit the ground state of the nonlinear Schr\"odinger equation is determined by the parabolic part of the nonlinear potential near its minimum, so its wave function approaches that of the 2D harmonic oscillator. Remarkably, the expression for the mass correction at $\alpha_\text{s}\zeta\gg1$ coincides with that obtained from the perturbation theory, Eq.~(\ref{eq:Sigma2=}).

Another variational function is the 2D analog of the one proposed in Ref.~\cite{Pekar1946}: 
\begin{equation}\label{eq:PsiExp=}
\Psi(\vec{r}_\|) = \frac{2\kappa}{3\sqrt\pi}\,(1+\kappa{r}_\|)e^{-\kappa{r}_\|},
\end{equation}
with the Fourier components
\begin{equation}
\Psi_\vec{k} = \frac{4\sqrt\pi/\kappa}{(1+k^2/\kappa^2)^{5/2}},\quad
(\Psi^2)_\vec{k} = \frac{16}3\,\frac{24+k^2/\kappa^2}{(4+k^2/\kappa^2)^{7/2}}.
\end{equation}
For this family, the energy functional and the correction to the effective mass can be evaluated analytically only at $z_0=0$:
\begin{subequations}\begin{align}
&\frac{F[\Psi(\vec{r}_\|)]}{\omega_\text{s}} = \int\frac{d^2\mathbf{k}}{(2\pi)^2}\left[k^2|\Psi_{\mathbf{k}}|^2 - \alpha_\text{s}\,\frac{2\pi}k\,|(\Psi^2)_{\mathbf{k}}|^2\right] \nonumber\\
& \qquad\quad
 = \frac{\kappa^2}3 - \frac{8\,575\,\pi}{36\,864}\,\alpha_\text{s}\kappa,\\
&\frac{\Delta{m}}m =  4\pi\alpha_\text{s}\int\frac{d^2\vec{k}}{(2\pi)^2}\,k|(\Psi^2)_{\mathbf{k}}|^2
= \frac{845\,\pi}{4\,608}\,\alpha_\text{s}\kappa^3.
\end{align}\end{subequations}
This gives $F_\text{min} = -\alpha_\text{s}^2\times0.401\ldots$, $\Delta{m}/m=\alpha_\text{s}^4\times0.759\ldots$.
For the Gaussian function the corresponding prefactors are $\pi/8=0.393\ldots$ and $\pi^2/16=0.617\ldots$, so the function~(\ref{eq:PsiExp=}) works slightly better at small $\alpha_\text{s}\zeta\ll1$.
The direct unconstrained minimization of the functional (see below) gives $0.405\ldots$ and $0.732\ldots$, respectively.

\subsection{Numerical minimization}
\label{app:LPnumerics}

Eq.~(\ref{eq:nonlinearSchroedinger}) is discretized on a uniform grid $s_n=nh$, $n=0,1,\ldots,n_\text{max}$, and the lowest eigenvector of the corresponding matrix is found on each iteration of the self-consistency loop. The differential operator is discretized in order to reproduce the integral
\begin{equation}
\int_0^\infty\left(\frac{df}{ds}\right)^2s\,ds\approx
\sum_{n=0}^\infty\left[\frac{f(s_{n+1})-f(s_n)}{h}\right]^2(n+1/2)h^2,
\end{equation}
so the matrix is tridiagonal, and numerical diagonalization is very efficient.
The step~$h$ and the upper cutoff $n_\text{max}h$ are chosen relative to the typical extent $1/\kappa_\text{G}$ of the Gaussian variational function~(\ref{eq:PsiGauss=}) minimizing the function~(\ref{eq:FGauss=}).

\begin{figure}
\includegraphics[width=0.48\textwidth]{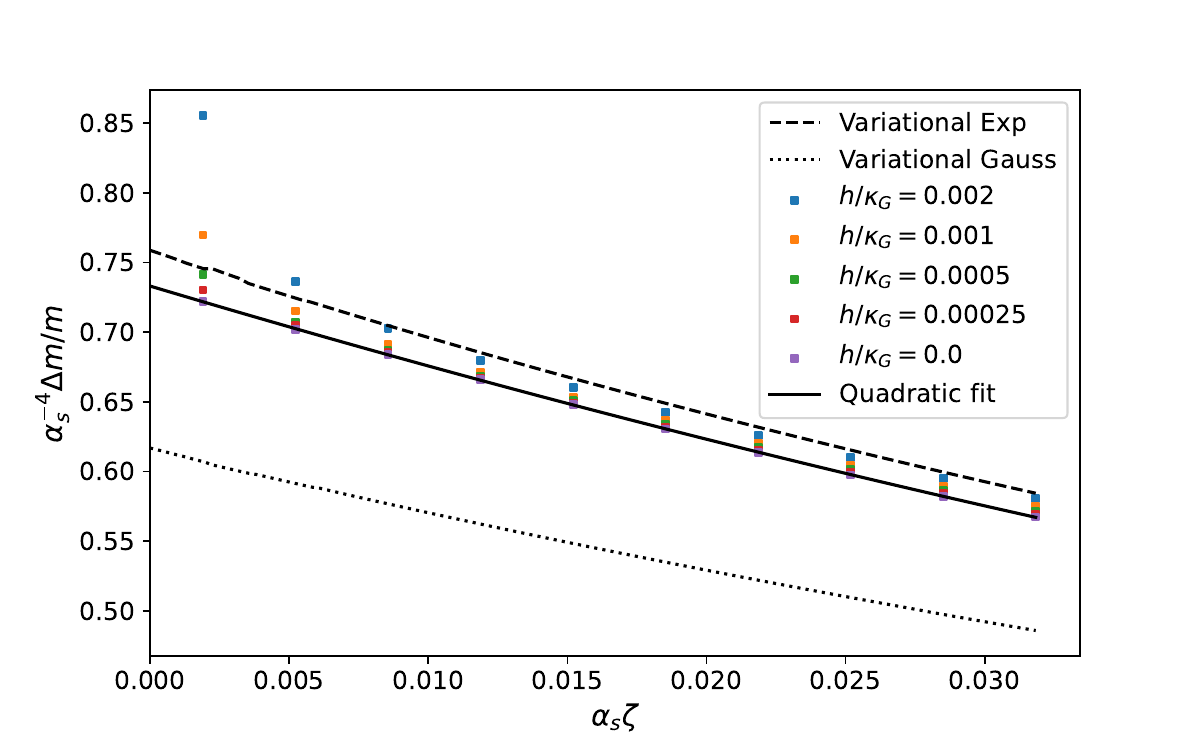}
\caption{\label{fig:LP_smallz}
Rescaled mass correction $\alpha_\text{s}^{-4}\Delta{m}/m$ for small values of the single dimensionless parameter $\alpha_\text{s}\zeta$ of the Landau-Pekar functional. The symbols correspond to different relative grid steps $h/\kappa_\text{G}^2=0.002,\,0.001,\,0.0005,\,0.00025$ and to their extrapolation to $h\to0$.
The grid cutoff $s_\text{max}=20/\kappa_\text{G}^2$.
The solid line shows a quadratic fit of the latter set, Eq.~(\ref{eq:dmm_fit}). The dashed and dotted lines show the result obtained from the two simple variational functions~(\ref{eq:PsiExp=}) and~(\ref{eq:PsiGauss=}), respectively.
}
\end{figure}

For small values of $\alpha_\text{s}\zeta$, rather small step sizes are required, so quadratic extrapolation to zero step was performed (see Fig.~\ref{fig:LP_smallz}), and the result was fitted to a quadratic function:
\begin{equation}\label{eq:dmm_fit}
\frac{\Delta{m}}m = \alpha_\text{s}^4 \left[0.733 - 5.97\,\alpha_\text{s}\zeta + 23.8\, (\alpha_\text{s}\zeta)^2\right].
\end{equation}
The mass correction at $\zeta\to0$ can be also obtained by setting $z_0=0$ from the very beginning. Then one can change variables as $s=2m\omega_\text{s}r_\|^2$, $\Psi(r_\|) = \sqrt{2m\omega_\text{s}}\,f_0(s)$; the resulting nonlinear Schr\"odinger equation has the same form as Eq.~(\ref{eq:nonlinearSchroedinger}), but one should replace $\zeta\to1$ everywhere, and 4 by 0 in both denominators in the integrand. Once the minimizer $f(s)$ has been found, the mass correction is given by a much simpler integral than Eq.~(\ref{eq:dmmE32E52}): 
\begin{align}
\frac{\Delta{m}}m= 
 {-16}\pi\alpha_\text{s}\int_0^\infty\frac{ds\,ds'}{\sqrt{s}+\sqrt{s'}}\,
K\!\left(\frac{4\sqrt{ss'}}{(\sqrt{s}+\sqrt{s'})^2}\right)
\nonumber\\
{}\times f^2(s')\,\frac{d}{ds}\,s\,\frac{df^2(s)}{ds}.
\end{align}
This calculation gives $\alpha_\text{s}^{-4}\Delta{m}/m = 0.732\ldots$.

\begin{figure}
\includegraphics[width=0.48\textwidth]{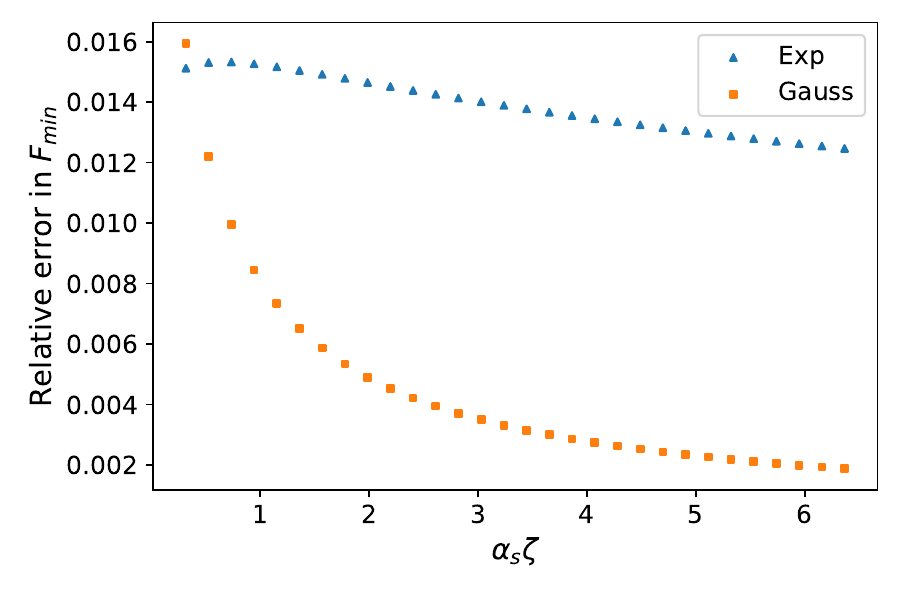}
\includegraphics[width=0.48\textwidth]{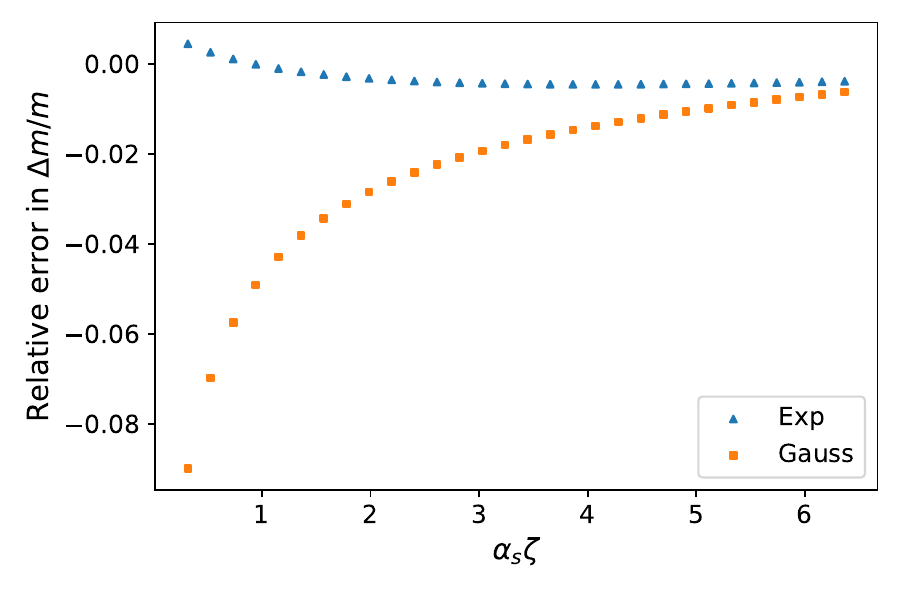}
\caption{\label{fig:variational}
Relative error in the functional minimum $F_\text{min}$ (upper panel) and in the mass correction $\Delta{m}/m$ (lower panel) for the two simple variational functions~(\ref{eq:PsiExp=}) and~(\ref{eq:PsiGauss=}) with respect to $|F_\text{min}|$ and $\Delta{m}/m$ found by the direct minimization on a grid.
}
\end{figure}

Finally, it is interesting to compare the results obtained from the two simple variational functions~(\ref{eq:PsiExp=}) and~(\ref{eq:PsiGauss=}) to the direct minimization in a wide range of~$\alpha_\text{s}\zeta$. The corresponding relative error for the functional minimum $F_\text{min}$ and the mass correction $\Delta{m}/m$ is shown in Fig.~\ref{fig:variational}. 
As expected, at large $\alpha_\text{s}\zeta$ the Gaussian wave function works better. However, even though it gives a significantly better energy already for $\alpha_\text{s}\zeta>1$, the mass correction is still given more precisely by the function~(\ref{eq:PsiExp=}) up to quite large $\alpha_\text{s}\zeta\sim7$.

\bibliography{cavityQED}

\end{document}